\documentclass[superscriptaddress,amsmath,amssymb,aps,pra, twocolumn]{revtex4-2}

\DeclareUnicodeCharacter{2032}{\ensuremath{\prime}}

\usepackage{graphicx}
\usepackage{dcolumn}
\usepackage{bm}
\usepackage[colorlinks]{hyperref}
\usepackage{float}
\usepackage{appendix}

\def\beq{\begin{equation}}
\def\eeq{\end{equation}}
\def\bsp{\begin{split}}
\def\esp{\end{split}}
\def\bea{\begin{eqnarray}}
\def\eea{\end{eqnarray}}

\begin{document}

\title{Survival acceleration}

\author{Yakov Bloch}
\affiliation{Raymond and Beverly Sackler School of Physics and Astronomy, Tel-Aviv University, IL-69978 Tel Aviv, Israel}

\author{Avshalom C. Elitzur}
\affiliation{Iyar, The Israeli Institute for Advanced Research, POB 651, Zichron Ya’akov 3095303, Israel}
\affiliation{Institute for Quantum Studies, Chapman University, Orange, California 92866, USA}

\begin{abstract}
Relativistic time dilation implies that an accelerating excited atom would have its lifetime prolonged in the lab frame. In this paper, we demonstrate a complementary effect: Longer-lived excited atoms turn out to have been accelerated. We propose the following experiment. An excited atom is prepared in a superposition of momenta. Due to relativistic time dilation, in the lab frame, the decay rate of components with higher momentum is diminished. As time passes without the atom emitting a photon, higher-momentum components become more probable. This results in a time-dependent shift of the expectation value of momentum. Taking into account first-order relativistic corrections, we calculate the force and acceleration undergone by the atom due to the fact that it did not emit a photon.             
\end{abstract}

\maketitle
Quantum mechanics is unique in that even events which did not occur, leave physical traces just because they \textit{could} have occurred. Reflecting the observer's knowledge about a quantum system, the wavefunction is modified when additional information is introduced. Such information may be provided, not only by a positive result (finding a particle at a certain position), but also by a negative one (not finding it). Pioneered by Renninger \cite{renninger1960messungen} and Dicke \cite{dicke1981interaction}, the study of such negative-result experiments, called interaction-free measurements, culminated in the work of Elitzur and Vaidman \cite{elitzur1993quantum}. Devising an interferometric thought experiment, they showed that information about a system can be gained without interacting with it. The study led to a variety of technological applications, ranging from sensing \cite{PhysRevApplied.19.054019} to imaging \cite{yang2023interaction} and noise characterization \cite{PhysRevA.110.032404}. 

IFM has led to further variants where apparent quantum non-events leave physical traces. In the Quantum Liar gedankenexperiment \cite{aharonov2018interaction}, two distant atoms become entangled even though the entangling photon exchange between them seem to have never taken place. Elitzur and Cohen \cite{elitzur2015quantum, cohen2015voices} have generalized this and many related quantum oddities as manifestations of the same "quantum oblivion" effect: A quantum  superposition is formed such that an interaction either occurs or not. Even in the latter case, physical consequences of the non-interaction can be demonstrated. 

The inception of the Elitzur-Vaidman effect was deeply rooted in relativistic thinking in that the click and non-click of two detectors measuring the position of a split particle are relativistically covariant. Nevertheless, the interplay of interaction-free measurement and relativistic effects, has not, to our knowledge, been explored. It is the aim of this paper to show that such interplay is a fertile ground for the prediction of novel effects.    

Consider an excited atom, prepared in a superposition of different momenta. Due to time dilation, these momenta give rise to different decay rates. Hence, the longer the atom does \textit{not} emit a photon, the more probable higher momenta become (since such momenta imply lower decay rates), resulting in a drift of the expectation value of momentum towards higher momenta. 

We work in the low-velocity regime of special relativity. This allows us to approximate time dilation effects without resorting to relativistic quantum wave equations. A fully relativistic treatment of the evolution of the wavefunction is not required in this limit, while the prediction of novel effects, such as the one described in what follows, becomes possible. 

We model the decay (or, rather, the non-decay) of the excited atom using a non-Hermitian effective Hamiltonian. This approach, describing a conditional evolution, is particularly common in the quantum jump and quantum trajectory theories \cite{RevModPhys.70.101, PhysRevB.111.064313, PhysRevB.110.094315, PhysRevA.101.062112, PhysRevA.103.052201, PhysRevLett.70.2273}. Under continuous observation, the evolution of a quantum system is separated into quantum jumps and no-jump evolution phases. Quantum jumps are discrete, stochastic transitions when a photon is detected while no-jump evolution captures the continuous, non-unitary evolution in the absence of detection, governed by
the effective Hamiltonian. To describe the proposed effect, in this paper, we only consider the non-unitary evolution when no emissions are detected. 

In what follows, we derive an expression for the relativistic acceleration of long-lived atoms prepared in a Gaussian superposition of momenta. We estimate the effect for a typical experimental setup. The work is concluded with a short discussion of the results.  

The proper lifetime of the excited atom in its rest frame is given by
\begin{equation}
    \tau_0 = \frac{1}{\Gamma_0},
    \label{tau}
\end{equation}
where $\Gamma_0$ is the proper decay rate. Due to time dilation, an atom moving with momentum $p$, and thus a Lorentz factor
\begin{equation}
    \gamma(p) = \frac{E(p)}{mc^2},
    \label{Lorentz}
\end{equation}
will have its lifetime extended in the lab frame. The decay rate in the lab frame, $\Gamma(p)$, is related to the proper decay rate $\Gamma_0$ by
\begin{equation}
    \Gamma(p) = \frac{\Gamma_0}{\gamma(p)}.
    \label{extended}
\end{equation}
Using the approximation for $\gamma(p)$ to first order in $\frac{1}{c^2}$
\begin{equation}
    \gamma(p) = \frac{E(p)}{mc^2} \approx \frac{mc^2 + \frac{p^2}{2m}}{mc^2} = 1 + \frac{p^2}{2m^2 c^2}.
    \label{approx}
\end{equation}
Therefore,
\begin{equation}
    \Gamma(p) = \frac{\Gamma_0}{1 + \frac{p^2}{2m^2 c^2}} \approx \Gamma_0 \left(1 - \frac{p^2}{2m^2 c^2} \right).
    \label{approx2}
\end{equation}
Let us consider an excited atom prepared in a Gaussian superposition of momenta
\begin{equation}
    \psi(p, 0) = \frac{1}{(2\pi \sigma^2)^{1/4}} \exp\left( -\frac{(p - p_0)^2}{4\sigma^2} \right)
    \label{momenta}
\end{equation}
We model the decay of the excited atom using the following non-Hermitian effective Hamiltonian
\begin{equation}
    \hat{H}_{\text{eff}} = \hat{H}_0 - i \frac{\hbar}{2}\hat{\Gamma}.
    \label{Ham}
\end{equation}
This Hamiltonian is valid when post-selecting on a null measurement outcome (i.e., the atom has not decayed by time $t$). The non-Hermitian term causes the norm of the wavefunction to decrease in time, reflecting the diminishing probability that the atom remains in the excited state. The non-Hermitian term encodes probability decrease due to decay, while the Hermitian part governs the motion of the atom, including a first-order relativistic correction. 

\begin{figure}[H]
    \centering
    \includegraphics[width=0.5\textwidth]{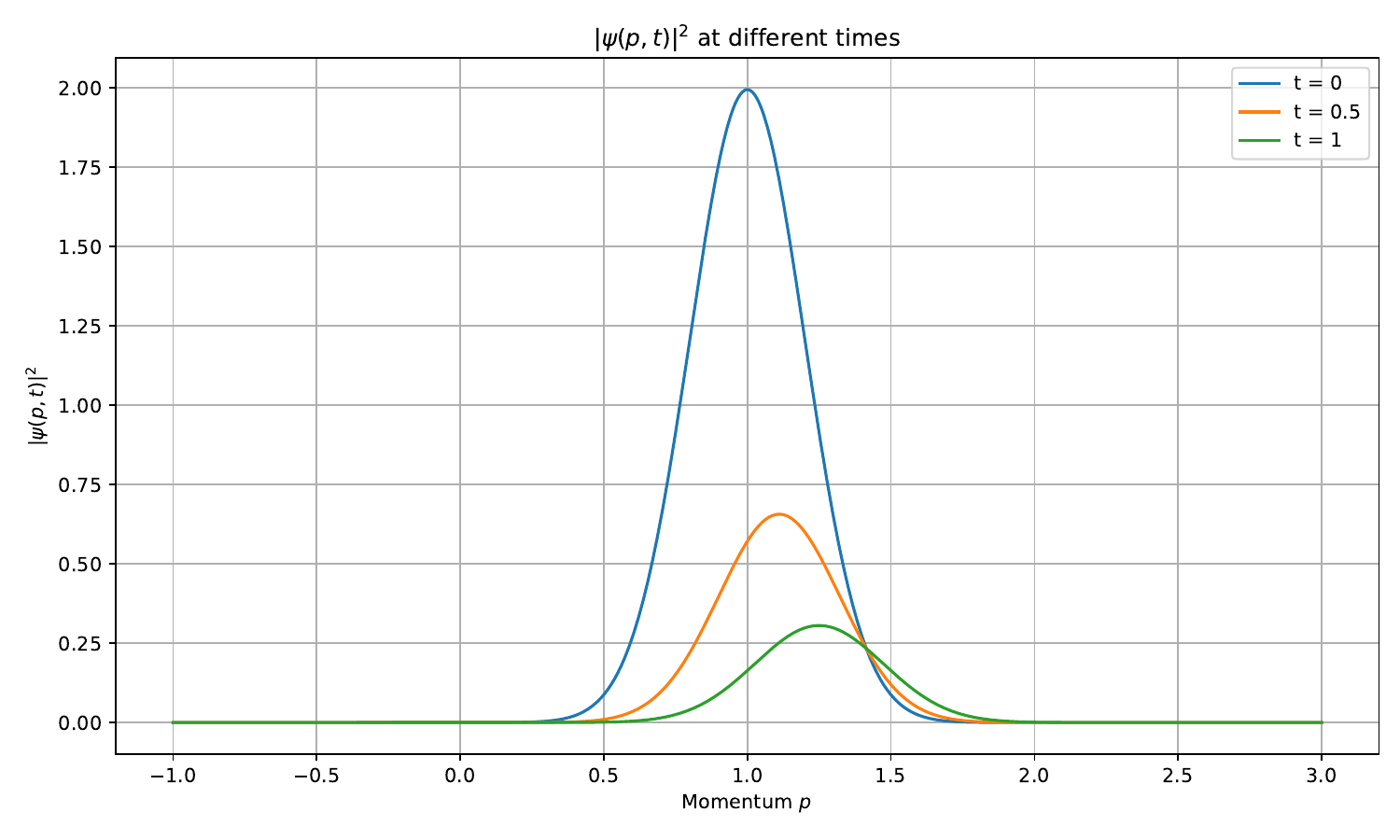}
    \caption{Non-normalized modulus squared of the wavefunction at different times, for $p_0 = 1$, $\sigma = 0.2$, $\Gamma_0 = 5$, $m=1$, $c=1$.}
    \label{fig:psi}
\end{figure}

To first order in $\frac{1}{c^2}$, we have
\begin{equation}
    E(p) \approx mc^2 + \frac{p^2}{2m} - \frac{p^4}{8m^3 c^2},
    \label{firstord}
\end{equation}
The Hamiltonian is, therefore,
\begin{equation}
    \hat{H}_{\text{eff}} = \frac{\hat{p}^2}{2m} - \frac{\hat{p}^4}{8m^3 c^2} - i \frac{\hbar}{2}\Gamma_0 \left(1 - \frac{\hat{p}^2}{2m^2 c^2} \right).
    \label{Ham2}
\end{equation}
The state evolves as
\begin{equation}
    |\psi(t)\rangle = e^{-i\frac{\hat{H}_{\text{eff}} t}{\hbar}}|\psi(0)\rangle,
    \label{evol}
\end{equation}
which results in
\begin{equation}
    \psi(p, t) = \psi(p, 0) \exp\left( -i \frac{E_0(p)}{\hbar} t \right) \exp\left( -\frac{1}{2} \Gamma(p) t \right).
    \label{evolp}
\end{equation}
First, let us calculate the expectation value of the momentum (conditioned on no decay) as a function of time. This is given by 
\begin{equation}
    \langle p\rangle_t = \frac{\int_{-\infty}^\infty p\,|\psi(p, t)|^2 dp}{\int_{-\infty}^\infty|\psi(p, t)|^2 dp}.
    \label{ptexp}
\end{equation}
The squared modulus of the (non-normalized) wavefunction is
\begin{equation}
    |\psi(p, t)|^2 = \frac{1}{\sqrt{2\pi \sigma^2}} \exp\left( -\frac{(p - p_0)^2}{2\sigma^2} + \frac{p^2 \Gamma_0 t}{2 m^2 c^2} \right)e^{-\Gamma_0 t}.
    \label{modsquare}
\end{equation}
Plotting the non-normalized wavefunction (figure (\ref{fig:psi})), we have
We see that the Gaussian mean is shifted towards higher momenta. It appears that the Gaussian shape is preserved. Let us calculate this shift. We rewrite the first exponent as
\begin{equation}
    -\frac{(p - p_0)^2}{2\sigma^2} + \frac{p^2 \Gamma_0 t}{2 m^2 c^2} = -\left( \frac{1}{2\sigma^2} - \frac{\Gamma_0 t}{2 m^2 c^2} \right)p^2 + \frac{p p_0}{\sigma^2} - \frac{p_0^2}{2\sigma^2}.
    \label{exp}
\end{equation}
The resulting wavefunction, therefore, remains Gaussian. Let us define the effective variance
\begin{equation}
    \frac{1}{2\sigma_t^2} = \frac{1}{2\sigma^2} - \frac{\Gamma_0 t}{2 m^2 c^2},
    \label{effvar}
\end{equation}
so that
\begin{equation}
    \sigma_t^2 = \left( \frac{1}{\sigma^2} - \frac{\Gamma_0 t}{m^2 c^2} \right)^{-1}.
    \label{effvar2}
\end{equation}
We see that at a certain threshold time the variance becomes imaginary. Due to the assumption that velocities are small compared to the speed of light, the expression is only valid for times significantly shorter than this threshold. Otherwise, the acceleration would push the momentum out of the regime where the approximation of the Lorentz factor, to first order in $\frac{1}{c^2}$, holds. Therefore, the derivation is only appropriate for times shorter than
\begin{equation}
    t \ll \frac{m^2 c^2}{\sigma^2 \Gamma_0}.
    \label{time}
\end{equation}
Completing the square, we write the shifted mean as
\begin{equation}
    \mu_t = \sigma_t^2\frac{p_0}{\sigma^2}.
    \label{shiftmean}
\end{equation}
The expectation value of momentum is simply the shifted mean of the Gaussian
\begin{equation}
    \langle p \rangle_t = \mu_t = \frac{p_0}{1 - \frac{\sigma^2 \Gamma_0 t}{m^2 c^2}}.
    \label{pt_final}
\end{equation}
To find the force, let us take the derivative of the expectation value with respect to time
\begin{equation}
    F(t) = \frac{d\langle p \rangle_t}{dt} = \frac{p_0 \sigma^2 \Gamma_0}{m^2 c^2 \left(1 - \frac{\sigma^2 \Gamma_0 t}{m^2 c^2}\right)^2}.
    \label{force}
\end{equation}
To first order in $\frac{1}{c^2}$, the force is constant and equal to
\begin{equation}
    F = \frac{p_0 \sigma^2 \Gamma_0}{m^2 c^2}.
    \label{force}
\end{equation}
We see that the force is proportional to the initial mean of the gaussian distribution and to the rest frame decay rate. When the initial expectation value of momentum is zero, the effect vanishes. Moreover, the larger the spread of the initial wavefunction and the smaller the mass of the atom, the larger the force. Conditioned on no photon emission, the atom experiences an apparent force as a consequence of relativistic time dilation. Non-emission favors the longer-lived higher-momentum components of the wavefunction, leading to an acceleration of the wavepacket. The effect analyzed is, essentially, a dynamical back-action of the measurement (more specifically, of the IFM type) on the system’s evolution.

The acceleration of an atom requires energy. The atom in our setup appears to accelerate without work being performed. Where does this energy come from? The answer to this question lies in the fact that the system is post-selected. The energy gained by the post-selected ensemble is compensated by the energy lost by the atoms that decayed early. The energy of the total ensemble remains constant. The fact that the energy of the post-selected ensemble is higher than the energy of the discarded ensemble is similar to the effect in a gedankenexperiment proposed by Aharonov et al. \cite{aharonov2021conservation}. A particle is prepared in a superoscillatory state within a box (superoscillations are functions that may locally oscillate faster than their fastest Fourier component \cite{berry2019roadmap}). Opening the box in the vicinity of the superoscillatory behavior, a high energy photon may emerge, the frequency of which arbitrarily (depending on the prepared wavefunction) exceeds the highest Fourier component of the prepared state. However, had the photon not emerged, the energy of the particle remaining in the box would be diminished to compensate for the extra energy of the photon in the post-selected ensemble, such that the energy of the total ensemble remains constant. The authors go on to describe the modular energy \cite{aharonov1969modular} exchange between the box and the particle. While such an analysis is beyond the scope of the present work, future work may refer to the modular energy exchange between the atom and the detector, which we briefly describe next.    

In what follows, we consider the entanglement of the atom with the detector, which we model as an immediate absorber. Suppose the detector (we call D) is an identical atom in a ground state in the proximity of the excited atom (we call S). The emission of a photon is therefore described by the process,
\begin{equation}
    |e\rangle_S|g\rangle_D \rightarrow|g\rangle_S|e\rangle_D,
    \label{process}
\end{equation} 
Where $e$ stands for excited and $g$ for ground. While the initial state of S and D is $|e\rangle_S|g\rangle_D$, at any later time $t$, the state becomes 
\begin{equation}
    |\Psi(t)\rangle_{SD} = \int_{-\infty}^{\infty}dp\left[\psi(p,t)|{e}\rangle_S|g\rangle_D + \chi(p,t)|g\rangle_S|e\rangle_D\right].
    \label{totstate}
\end{equation}
where $\psi(p,t)$ is the time-evolved momentum distribution (\ref{evolp}), and $\chi(p,t)$ is given by
\begin{equation}
    \chi(p, t) = \psi(p, 0) \cdot \sqrt{1 - e^{-\Gamma(p) t}} \cdot e^{-i \frac{E(p)}{\hbar} t}.
    \label{chi}
\end{equation}
And we see that the mere possibility of photon exchange between the two atoms leads to entanglement between system and detector. Frequent measurement of the detector leads, due to the Zeno effect \cite{PhysRevD.16.520}, to prolonged lifetimes for the system. This effect could be utilized in the experimental verification of the system's acceleration, since prolonged lifetimes amplify the momentum shift of the excited atom.  

Finally, let us estimate the magnitude of the force (\ref{force}), using realistic experimental parameters. We estimate the effect for a Rubidium-87 atom in an excited state. The values of the different constants are summarized in table (\ref{tab:parameters}). 
\begin{table}[h]
\centering
\begin{tabular}{|l|l|}
\hline
Parameter & Value \\
\hline
Atomic mass (\(m\)) & \(1.44 \times 10^{-25} \, \mathrm{kg}\) \\
Speed of light (\(c\)) & \(3.00 \times 10^8 \, \mathrm{m/s}\) \\
Excited state lifetime (\(\tau_0\)) & \(27 \, \mathrm{ns}\) \\
Decay rate (\(\Gamma_0 = 1/\tau_0\)) & \(3.70 \times 10^7 \, \mathrm{s^{-1}}\) \\
Initial mean momentum (\(p_0\)) & \(1.44 \times 10^{-27} \, \mathrm{kg \cdot m/s}\) \\
Initial velocity & \(\sim 1 \, \mathrm{cm/s}\) \\
Momentum spread (\(\sigma\)) & \(1.0 \times 10^{-28} \, \mathrm{kg \cdot m/s}\) \\
\hline
\end{tabular}
\caption{Experimental parameters for a Rubidium-87 atom in the 5P excited state.}
\label{tab:parameters}
\end{table}
\begin{equation}
    F = \frac{p_0 \sigma^2 \Gamma_0}{m^2 c^2}\approx 2.85 \times 10^{-43}\left[\mathrm{N}\right].
    \label{est}
\end{equation}
And the resulting acceleration of the atom is
\begin{equation}
    a = \frac{p_0 \sigma^2 \Gamma_0}{m^3 c^2}\approx 1.98 \times 10^{-18}\left[\frac{m}{s^2}\right].
    \label{est2}
\end{equation}
The validity of the first-order approximation of the Lorentz factor requires that the effective variance (\ref{effvar2}) remains real. For the chosen parameters (\ref{tab:parameters}), this yields the constraint
\begin{equation}
    t \ll \frac{m^2 c^2}{\sigma^2 \Gamma_0} \approx 5.04 \times 10^{15}[\mathrm{s}],
    \label{test}
\end{equation}
which is far greater than the proper excited state lifetime of Rubidium-87 ($\tau_0 \sim 27ns$). This confirms that the approximation holds for the experimental setup considered.

We conclude that while the effect is exceedingly small, it demonstrates a fundamental reciprocity in the interplay between quantum mechanics and relativity. While it is well known that an accelerated excited atom has, in the lab frame, its lifetime prolonged, we have shown that longer lived excited atoms are accelerated. Continuous negative-result observation pushes the wavefunction towards higher momenta (which become more probable) inducing a force acting on the atom and accelerating it. The acceleration of an atom requires energy. While the mean energy of the non-decayed ensemble is higher than the total ensemble mean, the extra energy comes at the expense of the discarded ensemble of decayed atoms such that the energy of the total ensemble is conserved. We have shown that the system and the detector become entangled even in the absence of photon emission, as was first noted in \cite{aharonov2018interaction}. 

We hope that the amplification of the effect (possibly via weak value amplification \cite{PhysRevX.4.011031}) will be addressed in future work. A measurement of the acceleration predicted in this paper would constitute an important validation of the results. A fully relativistic treatment of the problem may lead to an amplification of the effect in the relativistic regime of high atom velocities. Finally, we hope that this work inspires further investigations into the interplay of relativistic effects and interaction-free measurement.

\bibliography{citations}

\end{document}